\documentclass[reprint,
 amsmath,amssymb,aps,
prd,
]{revtex4-1}
\usepackage{graphicx,subfigure,color}
\usepackage{dcolumn}
\usepackage{bm}
\usepackage{hyperref}
\usepackage[mathlines]{lineno}

\newcommand{\be}{\begin{equation}}
\newcommand{\ee}{\end{equation}}
\newcommand{\bea}{\begin{eqnarray}}
\newcommand{\eea}{\end{eqnarray}}
\begin{document}

\title{\Large   Quantization  of the charge in Coulomb plus harmonic potential}
\author{Yoon-Seok Choun}
\author{Sang-Jin Sin}
\email{ychoun@gmail.com}
\email{sangjin.sin@gmail.com}
\affiliation{Department of Physics, Hanyang University, 222 Wangshimni-ro, Sungdong-gu, Seoul, 04763, South Korea }
\begin{abstract}
 We consider two models where the wave equation  can be reduced to the    effective Schr\"odinger equation whose   potential   contains both harmonic  and the Coulomb  terms, $\omega^{2}r^{2}-a/r$. The   equation   reduces to the biconfluent Heun's equation, and we find that the charge   as well as the energy  must be quantized and  state dependent.
  We also find   that two   quantum numbers are necessary  to count   radial degrees of freedom and suggest that this is a general feature of differential equation with higher singularity like the Heun's equation.
\end{abstract}
\date{\today}%



\maketitle

{\bf 1. Introduction: }
Since Schr\"odinger established the  equation in his name,
it has been believed that for any confining potential,
 there exists discrete energy levels although we may not write the analytic solution explicitly.
 However, recent experience \cite{Bag2019,Holo2019} told us that it may not be the case. When the potential has higher singularity, we need higher regularity condition. As a consequence,
 there is no normalizable solution unless  potential itself is quantized.

%

In this paper, we consider two models where the wave equation  can be reduced to the    effective Schr\"odinger equation whose   potential   contains both harmonic term $\omega^{2}r^{2}$ and the Coulomb term $-a/r$. The   equation of motion  reduces to the biconfluent Heun's equation, and we find that the charge   as well as the energy  must be quantized.
That is, both energy and charge must depends on the states.

We also find that due to the higher singularity, new quantum number appears. For example, in spherically symmetric   case,
apart from the radial quantum number $N$ and two angular ones $L,m$,
one more quantum number $K$ appears. It turns out that only when we combine two quantum numbers $N,K$, the full radial degree of freedom can be counted.
 We suggest that the presence of extra quantum numbers to count correct radial degrees of freedom is a general feature of differential equation with higher singularity like the Heun's equation.




 \vskip 0.3cm
{ \bf 2.  A quark model with  Coulomb and  linear   scalar potential}

Lichtenberg et.al\cite{Lich1982} found a semi-relativistic Hamiltonian  which leads to a Krolikowski type second order differential equation \cite{Krol1980,Krol1981,Todo1971} in order to calculate meson and baryon masses.
In the center-of-mass system, the total energy $H$ of two free particles of masses $m_1$,   $m_2$,   is
\begin{equation}
H = \sqrt{\vec{\mathbf{p}}^2 c^2 +m_1^2 c^4}+ \sqrt{\vec{\mathbf{p}}^2 c^2 +m_2^2 c^4}
\label{eq:ys}
\end{equation}
Let $S$ be  the Lorentz scalar interaction and $V$ be the interaction which  is a time component of a 4-vector. Then it is natural to incorporate the $V$ and $S$ into (\ref{eq:ys}) by making the replacements
\begin{equation}
H\rightarrow H+V, \hspace{1cm} m_i \rightarrow m_i+\frac{1}{2}S, 
\qquad i=1,2. \label{replace}
\end{equation}
We set $m_{1}=m_{2}=0$ and    introduce   $\quad V=- {a}/{r}$ and study its effect
for the spin-free Hamiltonian which was proposed   for  the meson ($q\bar{q}$) system in \cite{1985,1988,1991,2011}.
Then we have
\begin{footnotesize}
\begin{equation}
\left( E+\frac{a}{r} \right)^2 \psi (r) = 4\left[c^4\left(\frac{1}{2}br\right)^2 + c^2\left( P_r^2 + \frac{\hbar^2L(L+1)}{r^2}\right) \right]\psi (r)
\label{eq:67}
\end{equation}
\end{footnotesize}
where $b$ is a real positive constant and  we used  $\vec{\mathbf{p}}^2=  P_r^2 + \frac{\hbar ^2L(L+1)}{r^2}$ with $P_r^2 = -\hbar^2(\frac{\partial ^2}{\partial r^2} +\frac{2}{r} \frac{\partial}{\partial r} )$.
The linear scalar potential is for the confinement of the quarks bound by a QCD flux string with constant string tension $b$.
 Previously, we investigated the  model in the case  $V=0$ \cite{Bag2019} and concluded that
 for the consistency of the spectrum the current quark should have zero mass. Here   we want to introduce   $V=-a/r$  and   understand   its effect in the presence of the confining potential.

Factoring out the behavior near $r=0$ by  $\psi (r)= r^{\tilde{L}} f(r)$, above equation becomes
 \begin{equation}
 \frac{d^2 f( {r})}{d{ {r}}^2}  +  \frac{2(\tilde{L}+1)}{ {r}} \frac{d f( {r})}{d {r}} + \left(\frac{\mathcal{E}^2}{4} - \frac{b_0^2}{4}r^2  +  \frac{\mathcal{E}a_0  }{2r} \right) f( {r}) = 0 ,
\label{qq:4}
\end{equation}
where $ a_0 = {a}/{\hbar c}$,
$b_0 = {bc}/{\hbar }$,  $\mathcal{E}=  {E}/{\hbar c}$ and
 $
 \tilde{L}= -1/2+\sqrt{(L+1/2)^2- a_0 ^2/4}.
 $
If we further factor out the  near-$ \infty$ behavior by
$ f( {r}) = \exp\left( -\frac{b_0}{4} {r}^2 \right)y( {r})$  and
introduce   $\rho =\sqrt{b_0/2} r$, we get
\begin{equation}
\rho  \frac{d^2{y}}{d{\rho}^2} + \left( \mu \rho^2 + \varepsilon \rho + \nu  \right) \frac{d{y}}{d{\rho}} + \left( \Omega \rho + \beta \right) y = 0.
\label{eq:1}
\end{equation}
with
 $\mu=-2$, $\varepsilon=0$, $\nu=2(\tilde{L}+1) $, $ \beta = \epsilon a_0$
 and
 \be \Omega= \epsilon ^2   -(2\tilde{L}+3), \hbox{ with }
\epsilon = \mathcal{E}/\sqrt{2 b_0}. 
 \ee
  This equation is a  biconfluent Heun's  equation which
has a regular singularity at the origin and an irregular singularity  of rank  two  at the infinity\cite{Ronv1995,Slavy2000}.

Substituting $y(\rho)= \sum_{n=0}^{\infty } d_n \rho^{n}$ into (\ref{eq:1}), we obtain the   recurrence relation:
\bea
d_{n+1}&=&A_n \;d_n +B_n \;d_{n-1}  \quad
\hbox{  for  } n \geq 1, \label{eq:3} \hbox{ with } \\
 A_n&=&-\frac{\varepsilon n+\beta }{(n+1 )(n+\nu )},  \quad   B_n=-\frac{\Omega +\mu (n-1 )}{(n+1 )(n+\nu )}.\eea
For  $n=0$ term, only $d_{1},d_{0}$ appear and give    $d_1= A_0 d_0$.

Notice that when $a_0=0$, we have
\be A_{n}=\frac{\beta}{(n+1 )(n+\nu )}= \frac{\epsilon a_0}{(n+1 )(n+2(\tilde{L}+1) )} =0,\ee
 so that
the three term recurrence relation given in eq. (\ref{eq:3})
is reduced to two term recurrence relation between $d_{n+1}$ and $d_{n-1}$ and the Heun's equation is reduced to hypergeometric one.  That is, in this scaling, the Coulomb parameter is precisely the term increasing  the singularity order.
Similarly, if $b_{0}=0$, the system can also be mapped to a hypergeometric type.  The problem rises only when both potential terms are present.

Now, unless  $y(\rho)$ is a polynomial, $\psi (r)$ is divergent as $ \rho \rightarrow \infty$. Therefore we need to impose regularity conditions by which the solution is normalizable.
If we  impose two conditions \cite{Ronv1995,Slavy2000},
\begin{equation}
B_{N+1}= d_{N+1}=0\hspace{1cm}\mathrm{where}\;N\in \mathbb{N}_{0},
 \label{bb:2}
\end{equation}
the series expansion  becomes a polynomial of degree $N$:
as one can see from eq. (\ref{eq:3}),
eq. (\ref{bb:2}) is sufficient to give  $d_{N+2}=d_{N+3}=\cdots=0$ recursively. Then   the solution is a polynomial of order $N$,
$
  y_{N}(\rho)=  \sum_{i=0}^{N}d_{i}\rho^{i}.
 $
The question whether imposing both  equations in eq(\ref{bb:2}) is really necessary  for the finite   solution   was studied numerically and was concluded affirmatively in our earlier work \cite{Bag2019}.

In general, $d_{N+1}=0$ will define a $N+1$-th order polynomial  $ {\cal P}_{N+1} $  in $a_{0}$, so that Eq. (\ref{bb:2}) gives
 \be
\epsilon _{N,L}=\sqrt{2N+2\tilde{L}+3}, \quad
{\cal P}_{N+1}(a_{0})=0.  \label{Omega}
\ee
where the first   comes from $B_{N+1}=0$,  and it is nothing but the usual energy quantization condition.
Below we will examine the meaning of  the second condition by  constructing  explicitly the expressions of  a few low   order  polynomial $ {\cal P}_{N+1}$, which are given in the appendix.

One surprising fact is that for a given $N,L$,
there are many solutions which we can index by an integer $K$ which is smaller than $N$.
Depending on whether $N$ is even or odd, the distribution of solutions  of  $ {\cal P}_{N+1}(a_{0})=0$,  is  different.
For low lying $L$, the number of roots increases with $L$ but not regularly.  However, for $L\geq [N/2]-1$ the number of roots
is given by $[N/2]+2$. Here, $[x]$ is the integer part of $x>0$.
The presence of extra quantum number is natural from the algebraic point of view. But it is rather suprising from the counting degree of freedom.
We postpone the dynamics of associated $K$ to next section where we discuss the problem with a simpler model.

 Table~\ref{tb:2} shows   some  real roots of $ a_0^2 $'s for each $L$ with fixed $N=8$; here, 
 $a^{2}_{0i}$ is the
 $i$-th root of $a^{2}_{0}$ with given $N,L$.
  Similarly, Table~\ref{tb:3} shows  real roots of $ a_0^2 $'s for each $L$ when $N=9$.
\begin{table}[!htb]
\footnotesize
\begin{center}
\begin{tabular}{|l|l|l|l|l|l|} 
\hline
  & $a^{2  }_{00}$ & $a^{2  }_{01}$ &$a^{2  }_{02}$ & $a^{2  }_{03}$ & $a^{2  }_{04}$ \\
\hline
\hline
L=0 & 0 & none & none  & none & none \\ \hline
L=1 & 0 & 2.35525 & 7.90698 & none & none \\ \hline
L=2 & 0 & 2.97179 & 11.2403 & 21.9815 & none \\ \hline
L=3 & 0 & 3.43735 & 13.3617 & 28.3483 & 44.4635 \\ \hline
L=4 & 0 & 3.81341 & 14.9937  & 32.6448 & 54.7228 \\ \hline
L=5 & 0 & 4.12728 & 16.3243  & 35.9753 & 61.791 \\ \hline
\end{tabular}
\caption{ Roots of $ a_0^2 $  for $N=8$. }
\label{tb:2}
\end{center}
\end{table}
\begin{table}[!htb]
\scriptsize
\begin{center}
\begin{tabular}{|l|l|l|l|l|l|l|}
\hline
  & $a^{2  }_{00}$ & $a^{2  }_{01}$ &$a^{2  }_{02}$ & $a^{2  }_{03}$ & $a^{2  }_{04}$\\ \hline
\hline
L=0 & 0.374151 & none & none & none & none  \\ \hline
L=1 & 0.580422 & 4.80626 & none & none  & none  \\ \hline
L=2 & 0.71935 & 6.26714 & 16.0299  & 24.9066 & none \\ \hline
L=3 & 0.828203  & 7.32404 & 19.5432  & 35.397 & 48.5634   \\ \hline
L=4 & 0.917807 & 8.17078 & 22.1613  & 41.6303 & 63.7813 \\ \hline
L=5 & 0.993589 & 8.87727  & 24.2768  & 46.338 & 73.3714  \\ \hline
\end{tabular}
\caption{  Roots of $ a_0^2 $  for  $N=9$. }
\label{tb:3}
\end{center}
\end{table}
 \begin{figure}[!htb]
 \centering
  \subfigure[]
  { \includegraphics[width=0.47\linewidth]{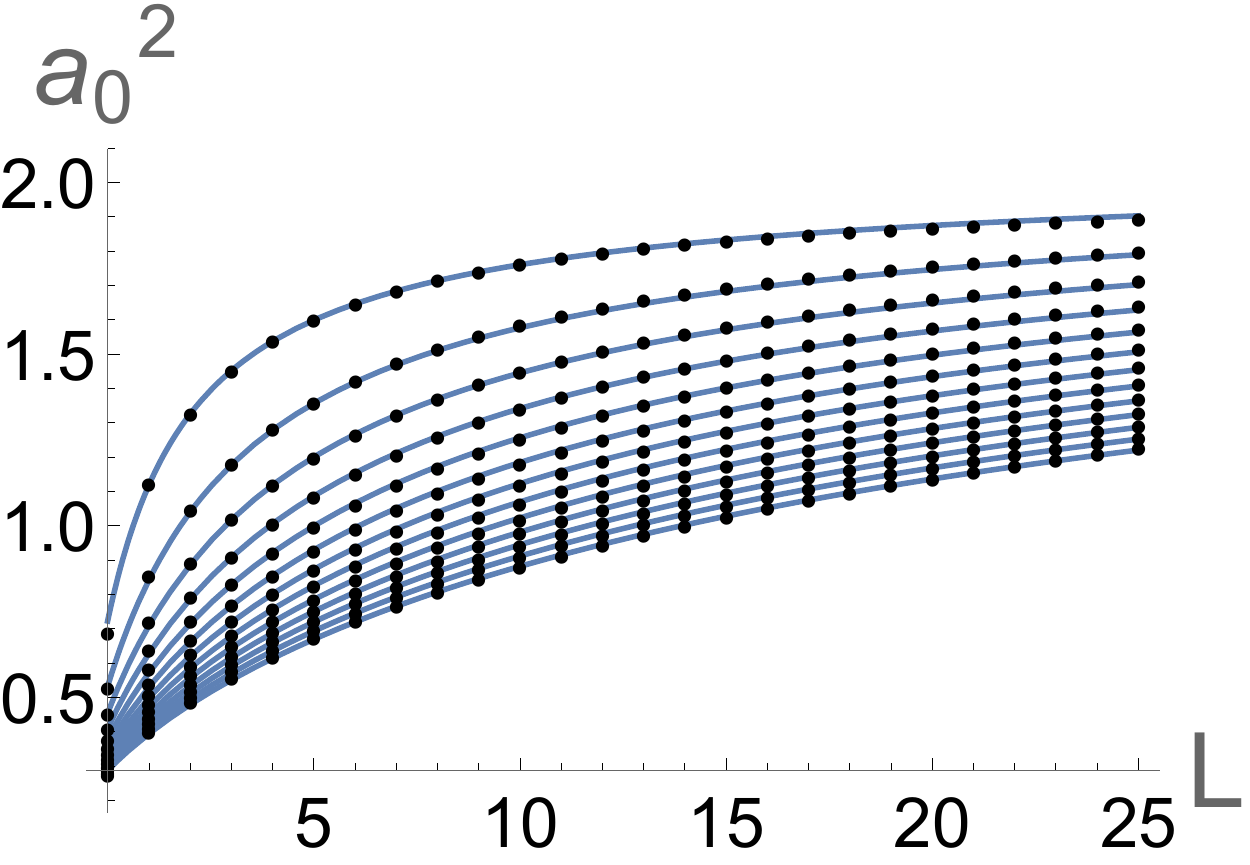}}
  \subfigure[]
 { \includegraphics[width=0.47\linewidth]{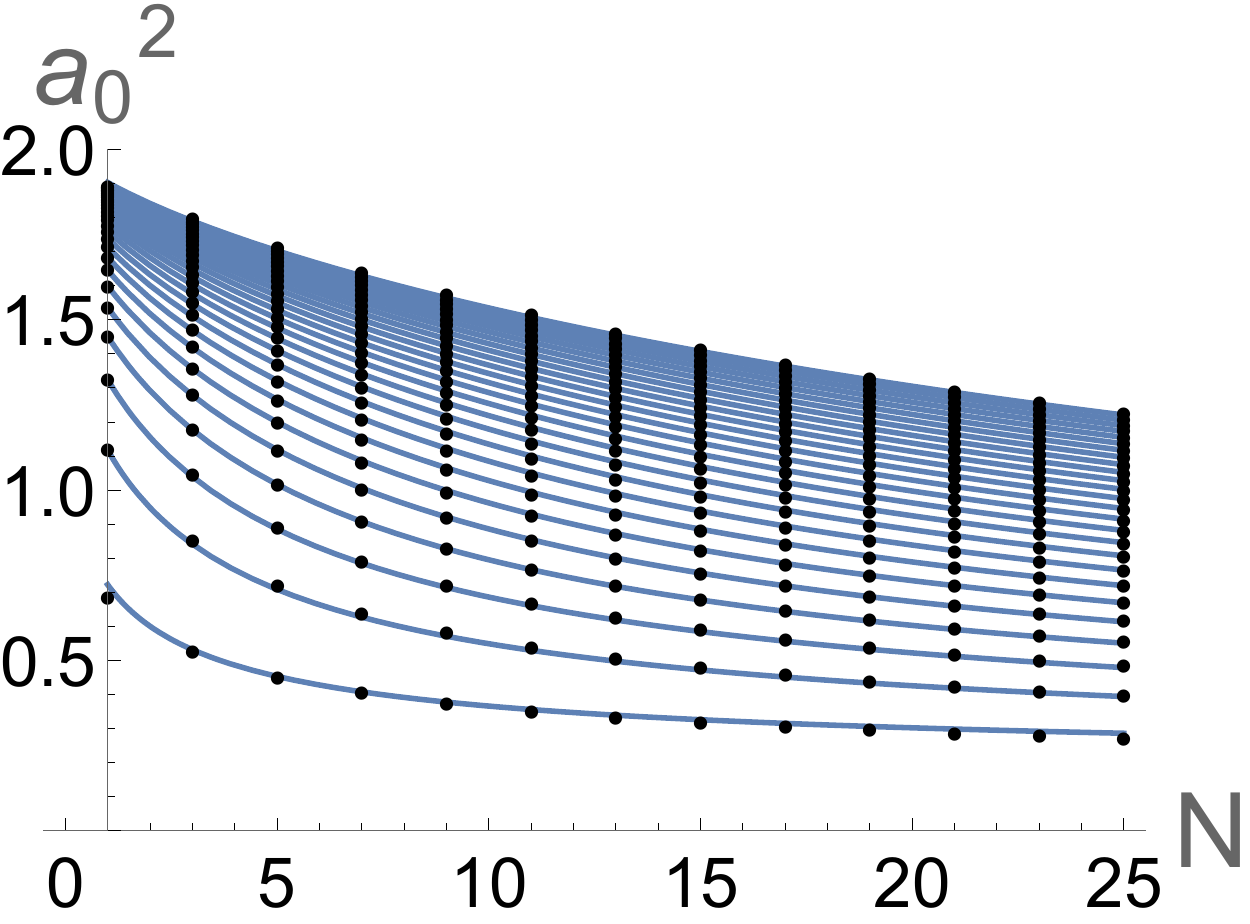}
}   \caption{ (a)     Fitting of $a_0^2$ data by eq.(\ref{bm1}), as functions of $L$.   The lowest   line is for $N=1$, the top  line  is for $N=25$.
(b)  Fitting of $a_0^2$ data  by eq.(\ref{bm1}) as functions of odd $N$ with $L$. the lowest   line is for $L=0$, the top one   is for $L=25$.
 In both figures (a)(b), N is odd.
}  \label{Root1}
\end{figure}

For lower value of $K$, we can find an approximate fitting function.
For example for $K=0$ and for odd $N$, it is given by
\be
a_{0,NL0}^2 \approx  1.22 \tan^{-1}\left( \frac{(L+1)^{0.6}-0.5}{0.55 N^{0.7}+0.5} \right) +0.18\label{bm1}.
\ee
We calculated  338 different values of $ a_0^2 $'s  at various $(N,L)$ and the result is the dots inFig.~\ref{Root1}.
These data fits well by above formula.
%
Notice also that for even $N$, $a_0=0$    is always a solution   for any  $L$.

By substituting  eq.(\ref{bm1}) into eq.(\ref{Omega}), we  can fit  the experimental data of   $E$, which is   the hadron mass.
  \be
    E_{ N,L} \simeq \sqrt{ {2\hbar c\cdot bc^{2}}
     \left[{2N+2+\sqrt{(2L+1)^2- a_{0,NLK} ^2} } \right] },  \label{bm5}
  \ee
  where   $a_0 = {a}/{\hbar c}$.
What is surprising is the fact that the charge parameter $a$
 should be quantized as values  approximately  given in  eq.(\ref{bm1}) if the charge is coming  in the presence of the linear scalar potential which gives the confinement.
Our treatment gives the analytic results in the presence
of the both linear potential together with Coulomb potential.
However, we must also comment that in the presence of the quark mass our method breaks down.

 \vskip.3cm
{ \bf 3.  Quantum dot    with  Coulomb   and    harmonic  potential }
Here we consider Non-relativistic Schr\"odinger equation with Coulomb potential and external harmonic oscillator potential   for a system of two electrons  in a three dimensional Euclidean space \cite{Caru2013,Reim2002,Siko1989,Merk1991}.
The Schr\"odinger equation is given by
\bea
&&\left[ -\frac{\hbar ^2}{2m}\left(\frac{d^{2}}{dr^{2}} +\frac{2}{r}\frac{d}{dr} \right) +V_{\mbox{eff}}(r)  \right]\psi(r) = E \psi(r), \\
&&\hbox{ with } V_{\mbox{eff}}(r)=\omega ^2 r^2 -\frac{a}{r}+ \frac{\hbar ^2}{2m} \frac{L(L+1)}{r^2} ,\label{heun}
\eea
Introducing   $\rho = r\left( \frac{2m \omega ^2}{\hbar ^2}\right)^{1/4}$,  above equation becomes
 \begin{equation}
\rho  \frac{d^2{\psi}}{d{\rho}^2} +\frac{2}{\rho} \frac{d{\psi}}{d{\rho}} + \left( \epsilon  -\rho^2  +\frac{a_0}{\rho} -\frac{L(L+1)}{\rho^2}\right) \psi = 0.
\label{jj:1}
\end{equation}
 where
 \be
 \epsilon=\frac{E}{\omega }\sqrt{\frac{2m}{\hbar ^2}},\quad   a_0=  \frac{a}{\sqrt{\omega}}\left(\frac{\hbar ^2}{2m }\right)^{-3/4}. \label{aa0}
 \ee
  Factoring out the behavior near $\rho =0$ by  $\psi(\rho)=\rho^{L} f(\rho)$,    it becomes  \begin{equation}
 \frac{d^2 f( {\rho})}{d{ {\rho}}^2}  +  \frac{2(L+1)}{ {\rho}} \frac{d f( {\rho})}{d {\rho}} + \left( \epsilon -\rho^2  +\frac{a_0}{\rho} \right) f( {\rho}) = 0.
\label{qq:4}
\end{equation}
 Factoring out    near   $ \infty$ behavior by
$ f( {\rho}) =e^{ - {\rho ^2}/{2}}y( {\rho})$,
we get  the standard form eq.(\ref{eq:1})
with
$$
 \mu=-2, \varepsilon=0, \nu=2(L+1),  \beta = a_0, \Omega= \epsilon -(2L+3).
$$
 Similarly, if we impose eq.(\ref{bb:2}), the series expansion  becomes a polynomial of degree $N$.  the solution becomes  a polynomial $
  y_{N}(\rho)=  \sum_{i=0}^{N}d_{i}\rho^{i}.
 $
In general, $d_{N+1}=0$ will define a $(N+1)$-th order polynomial  $ {\cal P}_{N+1} $  in $a_0$, so that Eq. (\ref{bb:2}) gives
 \be
\epsilon_{N,L}=2N+2L+3, \quad
{\cal P}_{N+1}(a_0)=0.  \label{Omega1}
\ee
where the first comes from $B_{N+1}=0$ which is the energy quantization condition.
Below we will examine the meaning of  the second equation. To do that we need explicit expressions of  a few lower order  polynomial $ {\cal P}_{N+1}$:
{\scriptsize
\bea
\label{app:10}
\begin{split} {\cal P}_{1}(a_0)&= a_0  ,\\
 {\cal P}_{2}(a_0)&=   a_0^2- 4(L+1)   ,\\
 {\cal P}_{3}(a_0)&=   a_0^3 - 4(4L+5) a_0  , \\
{\cal P}_{4}(a_0) & = a_0^4-20(2L+3) a_0^2  +144(L+1)(L+2) ,\\
{\cal P}_{5}(a_0) & = a_0^5-20(4L+7) a_0^3  +32(89+16L(2L+7))a_0 .
\end{split}
\eea
}
 In appendix, we   gave a few low order  polynomial $y_{N}(\rho)$ with $d_{0}=1$.

We  have seen  that  $a$ and $ \omega $ are related by eq. (\ref{aa0}) and  ${\cal P}_{N+1}(a_0)=0$ does not contain any dimensionful parameter.
This means that $a/\sqrt{\omega}$ should be a solution of a polynomial equation, which depends on $N,L$.
Such extra quantization is a consequence of the Heun's   equation.
For the hypergeometric equations,  the   recurrence relation is reduced to two term  after factoring out the asymptotic behavior.
There, we do not have $d_{N+1}=0$.
 Hence  to have a normalizable polynomial  solution,   we only need to fine tune  just one parameter, the energy,   For the Heun's equation,  we have to impose  two constraints, which in turn  request   the charge  quantization of the system.
In short,  its higher
   singularity  requests higher regularity condition.
This is the origin of the  charge quantization.

Notice that $a$ depends on the quantum numbers that parametrize quantum states.
  It means that when the electron make a transition from one state to  another, the charge parameter  must  be changed to a new value.
This raises the question,  how dynamics of one particle can change the potential energy which is
determined by the surrounding system.
In fact, $V$ is not the potential but the potential energy. The potential belongs to the surroundings
while the potential energy contains  both surrounding and particle information.  Therefore $a$ should be written as product of particle's charge $q$ times the charge $Q$ which makes the potential  $\phi_{Q}$, so that $V=q\phi_{Q}$.
When one say charge is quantized, what we mean is the quantization of  $q$.
In short, when the potential energy has higher singularity,  the charge as well as the energy should depends on the state. At first, this concept was rather drastic, but this is consequence of requesting $d_{N+1}$, whose necessity was confirmed by numerical investigation: without it, the shooting method did not work.

Notice that in this model,
 the energy  $\epsilon $  is linear in $N,L$ and does not depend on a quantized value of $ a_{0}^2$.
  Table~\ref{tb:4}  shows all roots of $ a_0^2 $'s for each $L$ for $N=4,5$.
 \begin{table}
\begin{tabular}{ cc }   
 $N=4$ &  $N=5$ \\
\scriptsize \begin{tabular}{|l|l|l|l|l|l|}
\hline
  & $a^{2 }_{00}$ & $a^{2 }_{01}$ &$a^{2}_{02}$     \\
\hline
\hline
L=0  &0 & 24.701 & 115.299   \\ \hline
L=1  &0  & 41.8531 & 178.147   \\ \hline
L=2  &0 & 58.414 & 241.586   \\ \hline
L=3  &0 & 74.7438 & 305.256    \\ \hline
L=4  &0 & 90.9604 & 369.04    \\ \hline
L=5  &0 & 107.114 & 432.886   \\ \hline
\end{tabular}  &  
\scriptsize \begin{tabular}{|l|l|l|l|l|l|l|}
\hline
  & $a^{2 }_{00}$ & $a^{2  }_{01}$ &$a^{2  }_{02}$   \\ \hline
\hline
L=0 & 6.38432 & 64.8131 & 208.803    \\ \hline
L=1 & 10.9664 & 102.965 & 306.069  \\ \hline
L=2 & 15.2359 & 140.155 & 404.609    \\ \hline
L=3 & 19.3928  & 176.898 & 503.709    \\ \hline
L=4 & 23.4959 & 213.403 & 603.101    \\ \hline
L=5 & 27.5688 & 249.768  & 702.664    \\ \hline
\end{tabular}  \\
\end{tabular}
 \caption{  Roots of $ a_0^2 $  }
 \label{tb:4}
 \end{table}

\begin{figure}[!htb]
 \centering
  \subfigure[]
  { \includegraphics[width=0.47\linewidth]{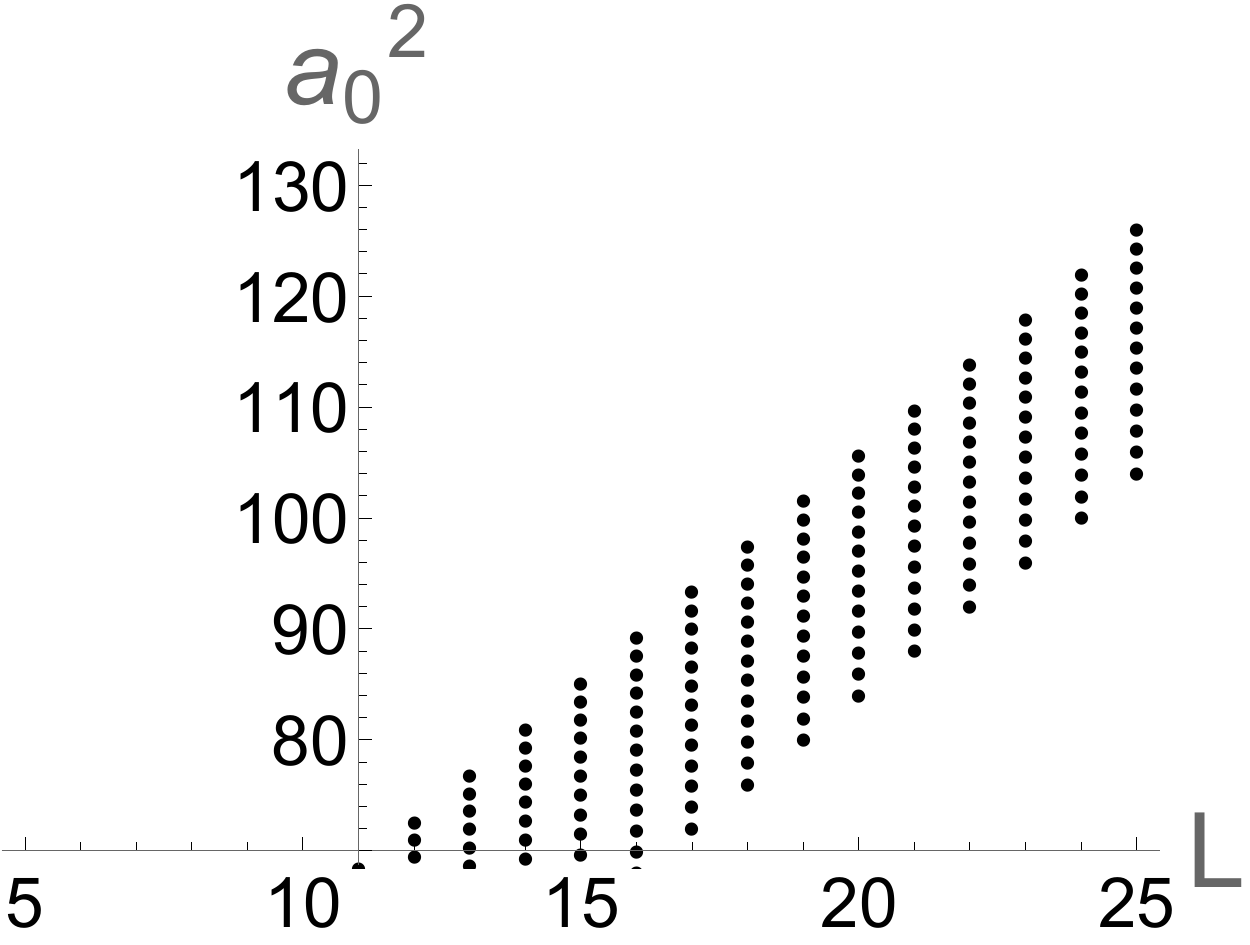}}
  \subfigure[]
 { \includegraphics[width=0.47\linewidth]{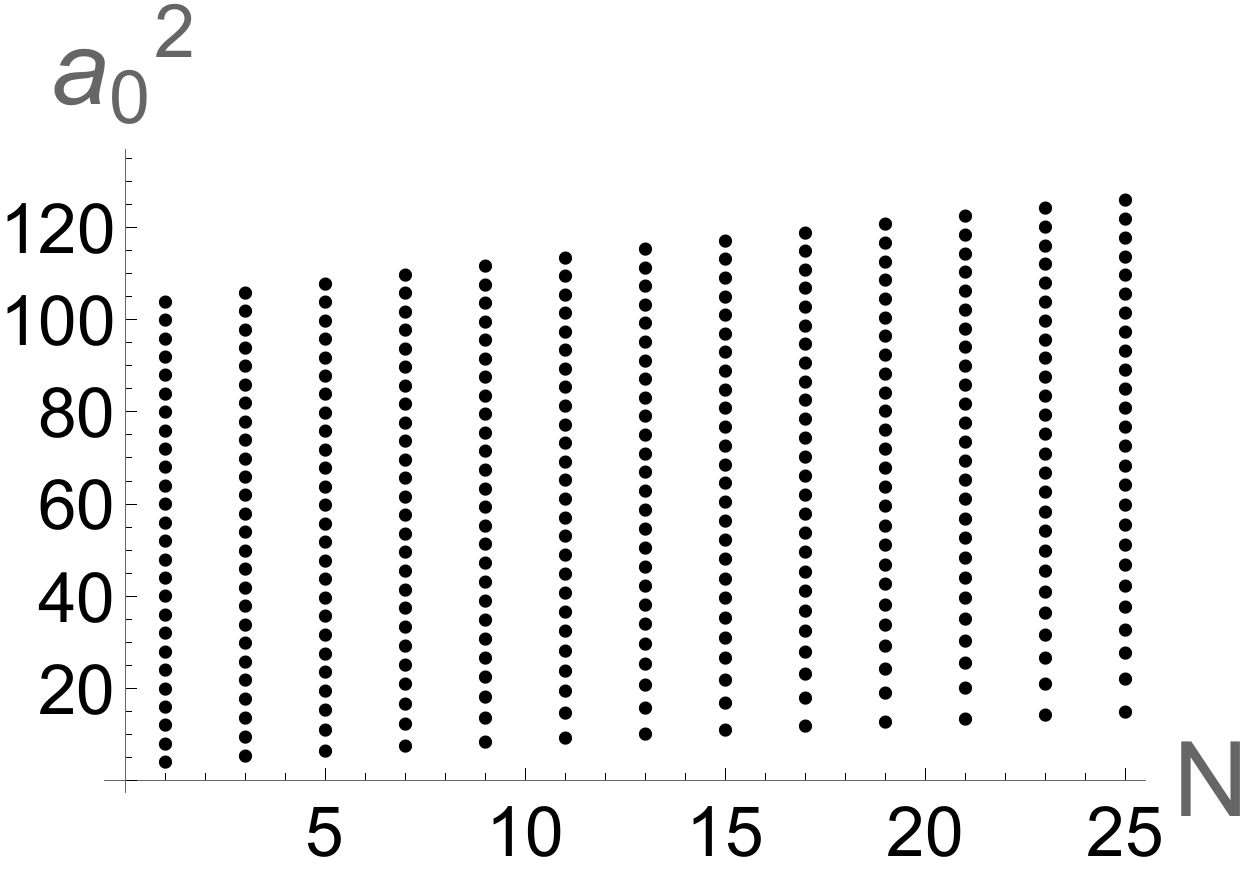}
}   \caption{  Roots of the smallest $a_0^2$, as function of L  and   N.
   In (a)  the lowest   line is for $N=1$, the top  line  is for $N=25$.  the lowest   line is for $L=0$, the top one   is for $L=25$.  In both figures (a)(b), N is odd.
}  \label{Root3}
\end{figure}

Since the quantized values of $a_{0}^{2}$ depends on three quantum number, we choose the  $K=0$ sector of $a_0^2$ with given $(N,L)$.
Then, Fig.~\ref{Root3} shows us that   $a_0^2$ is roughly linear  in $N,L$
for odd $N$.  For even $N$, the $K=0$ sector gives $a_{0}^{2}=0$.
%

For the figure, we calculated  338 different values of $ a_0^2 $'s  at various $(N,L)$.
 From the explicit calculation,  we find the following pattern: List N+1 $ a_0^2$ in the increasing order such that $a_{0,K}$ is $K$-th one, $K=0,1,\cdots ,\left[ N/2\right]-1$: Here,  $[x]$ means interger part of the positive real number $x$.
 Then  although the total number of nodes is $N$, some of them are in the negative region of $\rho$.
  The polynomial  with   $ a_{0K} $ has
 $N-\left \lfloor{N/2}\right \rfloor+K$ nodes in the region $\rho>0$.
 Therefore $K$  counts the number of nodes that crossed $\rho=0$ compared with $K=0$ in the positive domain.
 \begin{figure}[!htb]
 \subfigure[$K=0$, $L=0$]
 { \includegraphics[width=0.3\linewidth]{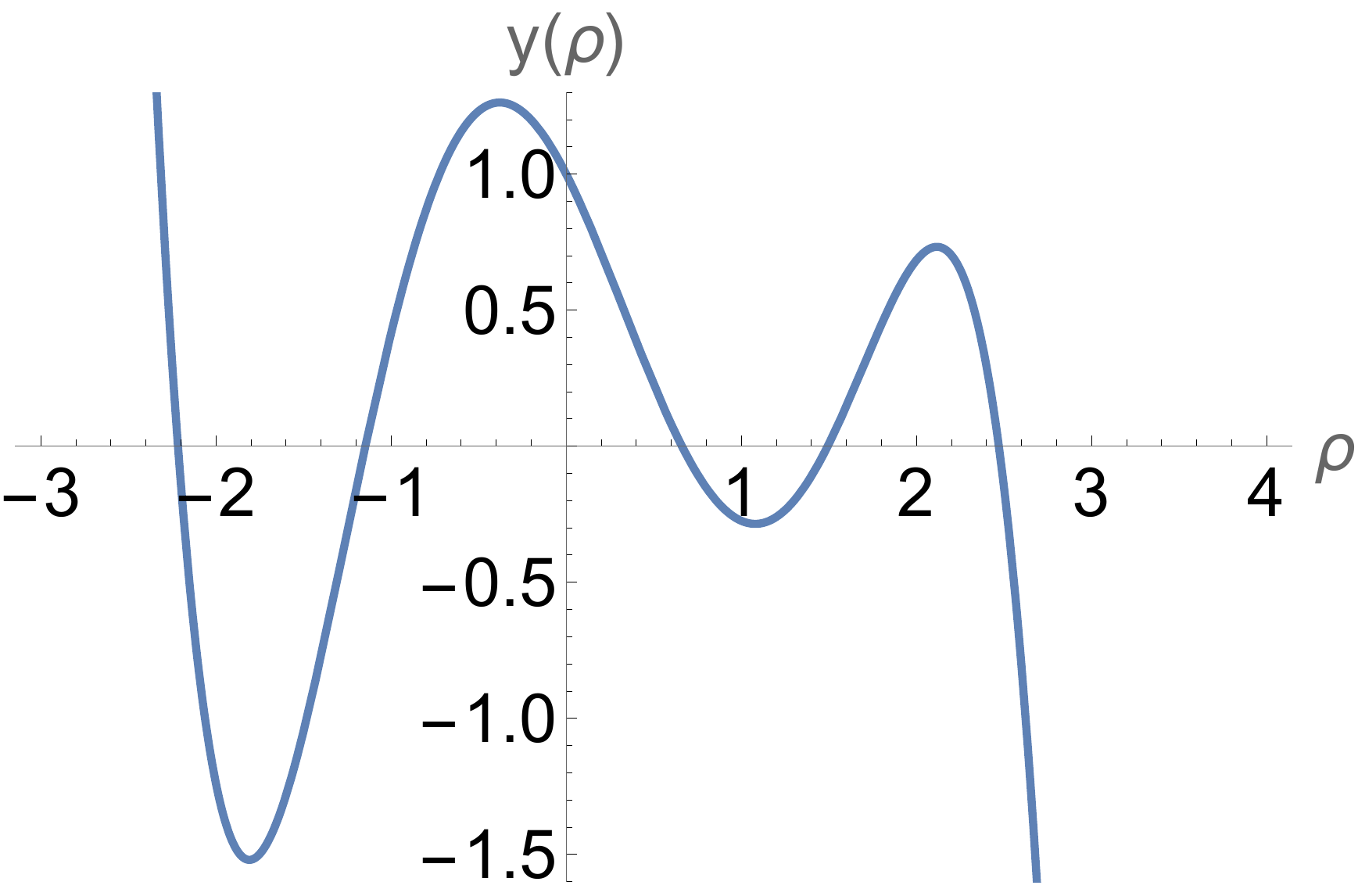}
 }
   \subfigure[ $K=1$, $L=0$]
  { \includegraphics[width=0.3\linewidth]{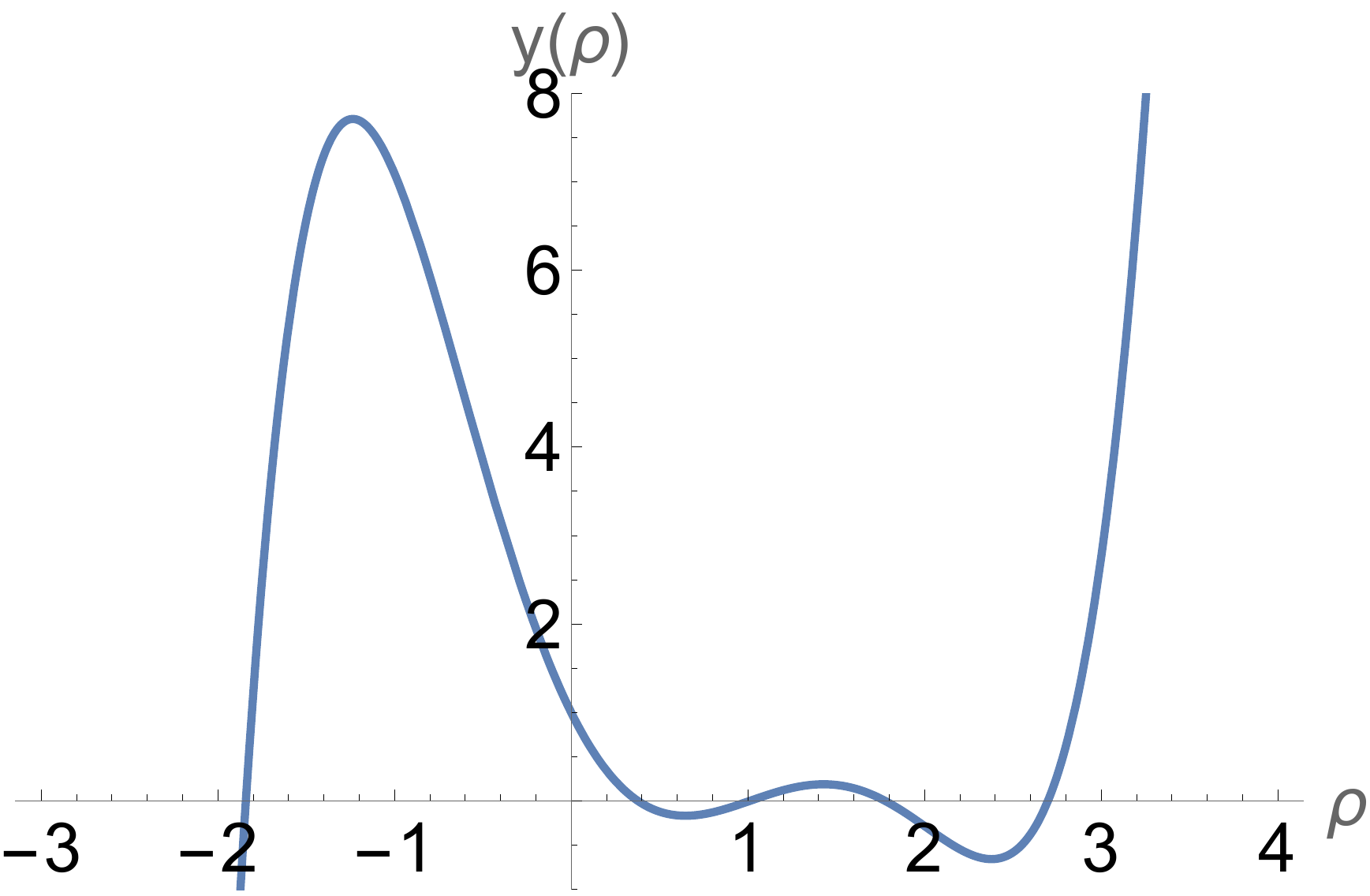}
  }
  \subfigure[$K=2$, $L=0$]
 { \includegraphics[width=0.3\linewidth]{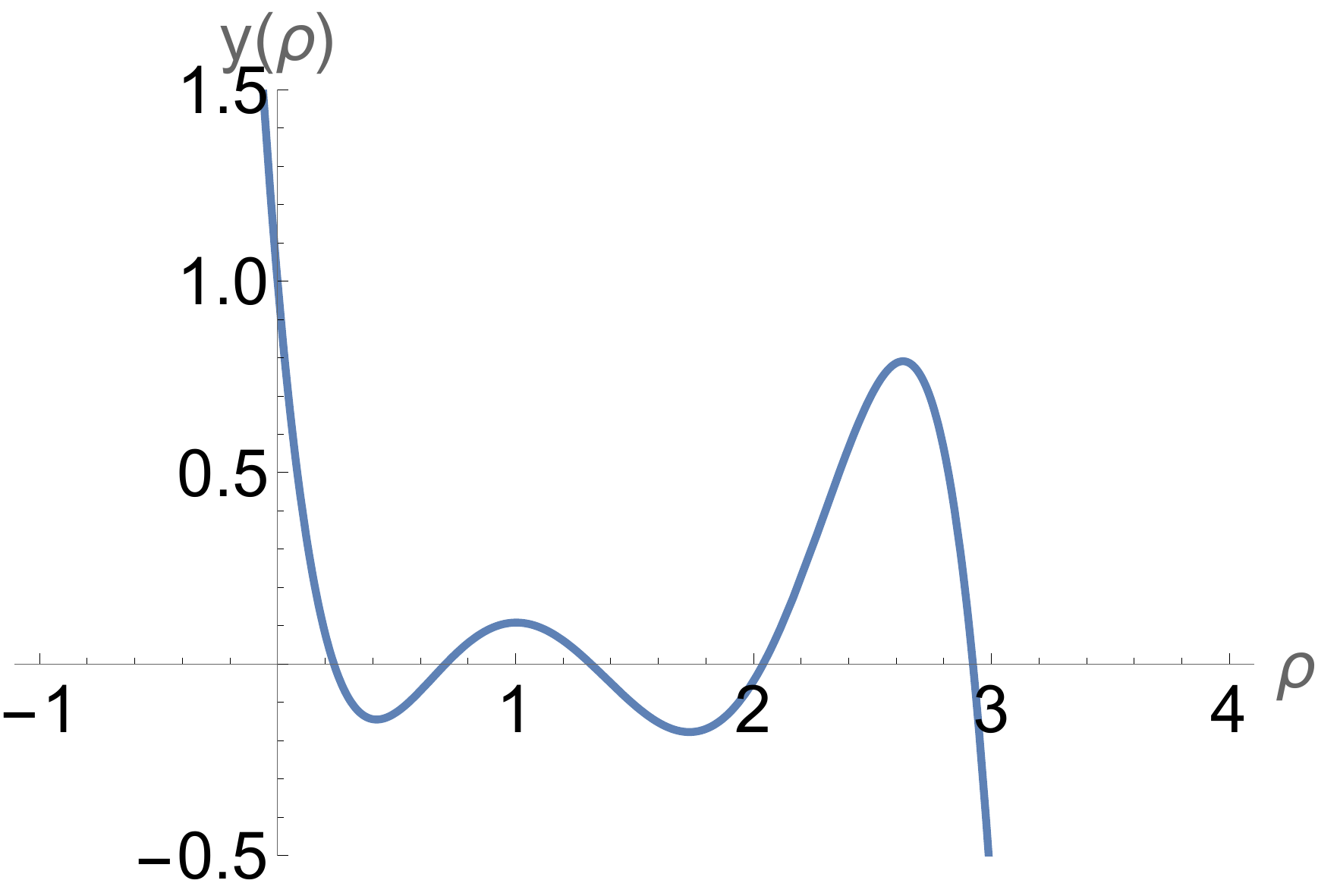}
 }
  \caption{ $y_{5}$ for  various  $ a_{0K} $ with $K=0,1,2$  and $L=0$.  For each $ a_{0K}$,   the number of positive roots is given by  $N-\left \lfloor{N/2}\right \rfloor+K=K+3$.}
\label{node4}
\end{figure}

In three dimension, spinless Hydrogen atom has three quantum number: $N,L,m$: $N$ for radial and the other two for angular momentum.
However, in the presence of the harmonic potential,  the charge and energy have discrete values depending on four quantum numbers $N,L,m,K$, which shows apparent mismatch between the number of degrees of freedom  and  that of quantum numbers.
However, as we have shown above,  with $K=0$, only half of the nodes of radial wave function are on positive region. This means that
the radial solution for fixed $K$, say $K=0$, can not span arbitrary shape of radial function in the positive region.   In fact,  $K$ counts the number of nodes which is moved from negative  to positve region compared with $K=0$ case.
This means  that $N$ together with $K$ counts full radial degrees of freedom, and without the extra quantum number $K$,  the solutions can not be a basis of  the   radial wave functions.

We expect that the presence of extra quantum number to count correct radial degrees of freedom is a general feature of differential equation with higher singularity like the Heun's equation.

{\bf 4.     Discussion}:
Caruso et.al\cite{Caru2013} investigated non-relativistic 2-D radial Schr\"odinger equation which can be related to ours just by shifting $L$ to  $L+1/2$ in (\ref{heun}).   They obtained  part of result of section 3 of this paper but they interpreted the result as the quantization of
  $\omega$, the coefficient of the harmonic potential.
The quantization of  $\omega $  would imply   that the single particle dynamics changes the  potential's parameter,  which does not sound plausible.  In our case, $a$ is split into particle charge
$q$  and charge $Q$ in the potential, so that Couomb term can be written as $V_{Coulomb}=q\phi_{Q}(r)$.  $q$  is a property of the particle, therefore
dependence of the particle charge on the state is natural although the concept is still not familiar so far. 
In field theory, charge depends on probe energy scale due to the renormalization.
So the state dependence of the charge can be regarded as discrete renormalization of the charge  induced by  smoothing out process of the the singularity of the potential. 

\section*{Acknowledgements}
 This  work is supported by Mid-career Researcher Program through the National Research Foundation of Korea grant No. NRF-2016R1A2B3007687.

\onecolumngrid
\appendix
\section{${\cal P}_{N}$ for  $N=0,1,...,5$}
{\scriptsize
\bea
\label{app:1}
\begin{split} {\cal P}_{1}(a_{0})&=  a_0 ,\\
 {\cal P}_{2}(a_{0})&=  a_0^2 \left( 4+\sqrt{(2L+1)^2-a_0^2}\right) -2\left( 1+\sqrt{(2L+1)^2-a_0^2}\right)  ,\\
 {\cal P}_{3}(a_{0})&=  a_0 \left( -a_0^2 \left(6+\sqrt{(2L+1)^2-a_0^2}\right)+12+8\sqrt{(2L+1)^2-a_0^2}\right)  , \\
{\cal P}_{4}(a_{0}) & =  -a_0^6 +a_0^4 \left( 85+4L(L+1)+16\sqrt{(2L+1)^2-a_0^2}\right) \\
 & -8 a_0^2 \left( 47+10L(L+1) +25\sqrt{(2L+1)^2-a_0^2}\right) \\
 &+144\left(L^2+L+1+\sqrt{(2L+1)^2-a_0^2} \right)
\end{split}
\eea
}

\section{ $ y_{N}(\rho )$ polynomials for $N=0,1,...,5$}
We lists expressions of  a few lower     order  polynomial $ y_{N}(\rho )$:
{\scriptsize
\bea
\label{app:100}
\begin{split} y_{0}(\rho )&=  1,\\
 y_{1}(\rho )&=    1 - \frac{a_0 \rho }{2(1 + L)} ,\\
 y_{2}(\rho )&= 1 +\frac{(a_0^2-8(L+1))\rho ^2 -2a_0(2L+3)\rho }{4(L+1)(2L+3)} \\
 y_{3}(\rho ) & =  1+\frac{a_0(36+28L-a_0^2)\rho ^3 +6(L+2)(a_0^2-12(L+1))\rho ^2 -12(L+2)(2L+3)a_0 \rho }{24(L+1)(L+2)(2L+3)}\\
  y_{4}(\rho ) & =  1+\frac{1}{96(L+1)(L+2)(2L+3)(2L+5)} \Bigg\{ (384(L+1)(L+2)-4(25+16L)a_0^2+a_0^4)\rho^4 +4(2L+5)(52+40L-a_0^2)a_0 \rho^3  \\
&-24(L+2)(2L+5)(16+16L-a_0^2)\rho^2 -48(L+2)(2L+3)(2L+5)a_0\rho \Bigg\} \\
   y_{5}(\rho ) & =  1+\frac{1}{960(L+1)(L+2)(L+3)(2L+3)(2L+5)} \Bigg\{ -a_0(6880+2384L^2+24L(354-5a_0^2)+a_0^2(a_0^2-220))\rho^5 \\
&+10(L+3)(720(L+1)(L+2)-4(35+22L)a_0^2+a_0^4)\rho^4 \\
&+40 a_0(L+3)(2L+5)(68+52L-a_0^2)\rho^3
-240(L+2)(L+3)(2L+5)(20+20L-a_0^2)\rho^2 \\
&-480 a_0(L+2)(L+3)(2L+3)(2L+5)\rho \Bigg\}
\end{split}
\eea
}

\end{document}